\documentclass[aip,rsi,reprint,graphicx]{revtex4-1} % for checking your page length
\usepackage[english]{babel}
\usepackage{textcomp}			% for different mathematical symblos 
\usepackage{subscript}		% for text formating like subscript
\usepackage{graphicx}			% for figures
\usepackage{array}				% for tables

\begin{document}

\title{A scalable, fast and multichannel arbitrary waveform generator}

\author{M.~T.~Baig}
\author{M.~Johanning}
\author{A.~Wiese}
\author{S.~Heidbrink}
\author{M.~Ziolkowski}
\author{C.~Wunderlich}
\email{wunderlich@physik.uni-siegen.de}
\affiliation{Department of Physics, University of Siegen, Germany} 

\date{\today} % It is always \today, today, but any date may be explicitly specified

\begin{abstract}
This article reports on the development of a multichannel arbitrary waveform generator (MAWG) that 
simultaneously generates arbitrary voltage waveforms on 24 independent channels with 
a dynamic update rate of up to 25\,Msps. A real-time execution of a single waveform and/or
sequence of multiple waveforms in succession, with a user programmable arbitrary sequence 
order is provided under the control of a stand-alone sequencer circuit implemented using an 
FPGA. The device is operated using an internal clock and can be synced to other devices by 
means of  transistor–transistor logic (TTL) pulses. The device can provide up to 24 independent voltages in the range of up to \textpm{}\,9\,V with a dynamic update-rate of  up to 25\,Msps and a power consumption of less than 35\,W. Every channel can be  programmed for 16 independent arbitrary waveforms that can be accessed during run time with a minimum switching delay of 160\,ns. The device has a low-noise of 250\,\textmu{}V\textsubscript{rms} and provides a stable long-term operation with 
a drift rate below 10\,\textmu{}V/min and a maximum deviation less than 
\textpm{}\,300~\textmu{}V\textsubscript{pp} over a period of two hours. \end{abstract}

%\keywords{waveform generator, signal generator, digital to analog converter,
%versatile voltages generator, fast updating DC-voltages generator.}

\maketitle

\section{Motivation}

Ion traps have promising potential in
the field of quantum information science \citep{Cirac_1995,Blatt_2008,Johanning_2009,Blatt_2012},
and in particular micro-structured Paul traps are well suited for this purpose. They can have multiple zones for 
trapping, processing and storing of atomic ions \citep{Kielpinski_2002,Schulz_2008,Kaufmann_2012,Wilpers_2012,Kunert_2013}. Scaling such traps up for 
quantum information processing requires a large number of independent high bandwidth, 
low noise voltage signals. 

In this approach, a register of quantum bits (qubits) is stored in the internal states 
of laser-cooled trapped ions (forming a Wigner crystal), and in some implementations
in motional states of their normal modes \citep{Cirac_1995}. 
Wigner crystals of increasing length become difficult to cool and
to protect against environmental influences and thus are subject to
decoherence. This is detrimental for tasks that rely on the computational power of quantum superpositions and
entanglement. Splitting the quantum register into crystals of manageable
size and entangling them on demand in a quantum network is regarded
as a promising and straight-forward solution. This can be achieved
in segmented ion traps where storage, shuttling and processing zones
\citep{Wineland_1998,Kielpinski_2002} are realized by a large number of DC electrodes.
These type of traps are often regarded as a prerequisite to scalable
quantum information processing using trapped ions.

Specific implementations of such a partitioned quantum register require
shuttling of ions in order to be able to exchange information between
independent crystals and this is ideally done with low heating of
the motional state \citep{Wineland_1998}. This can be achieved by
shuttling of the ions in an adiabatic fashion, at the expense of long
shuttling times \citep{Rowe_2002}, or in an tailored diabatic way
which can be optimized for low heating \citep{Schulz_2008, Walther_2012, Bowler_2012}. The latter
requires voltages with a fast update rate.

Another building block of this type of quantum information processing
is splitting and merging strings of ions, again requiring precise
control over a large number of voltages \citep{Eble_2010}. Various
proposals for processing quantum information exist, among others the
network model \citep{Steane_1998,Sasura_2002}, or measurement based
quantum computation \citep{Raussendorf_2003}, relying on highly entangled
graph states, which could be generated in ion traps \citep{Wunderlich_H_2009}.

Shuttling and the tailoring of interactions cannot only be done by
manipulating the DC voltage, but also the RF voltage used for radial
trapping of an ion string \citep{Kumph_2011}. A general purpose quantum
processor would be a field programmable trap array (FPTA). 
Independent arbitrary time-dependent potentials (e.g., sums of dc and rf voltages) can be applied to each  element of the FPTA, allowing for the 
creation of arbitrary trapping potentials (compared to designs optimized 
for a fixed lattice \citep{Schmied_2009}) and the investigation of various 
lattice types and quantum simulations thereof with a single programmable trap (a similar concept is discussed in ref.\citep{Lybarger_2010_PhD}). 
For certain types of ions and trap sizes the device presented here already fulfills all
the necessary requirements. This type of application might require -- depending on the 
selected ion and trap design -- higher amplitudes or update rates than supplied by the 
device described here, and, to reduce heating, narrow band filtering of the RF component.  

Similarly, quantum simulators require a high degree of control to simulate the 
features of other quantum systems \citep{Johanning_2009,Schneider_2012,Blatt_2012}. 
The interaction between trapped ions could be tailored to mimic an entirely
different quantum system, for example by shaping the axial trapping
potential \citep{Wunderlich_2003,Khromova_2012}. This application, too, requires
a large number of independent voltages.

Furthermore, in the field of experimental quantum optics, laser fields are
often used to manipulate the internal or motional state of atoms,
and need to be shaped in time and space. Shaping the temporal profile
of laser pulses can be realized using acousto-optic modulators (AOMs),
where control signals can be used to generate arbitrary amplitude
and frequency patterns of a laser \citep{DeMaria_1965,Jolly_2004}.
The spatial profile of a laser beam can be manipulated by a spatial
light modulator (SLM), typically implemented as a liquid crystal display
(LCD) or digital micro-mirror devices. The SLMs often take digital
signals for control using pulse width modulation, but analog implementations 
exist \citep{Serati_1995} potentially requiring again a large set
of independent voltages. In the same fashion the spectral and accordingly
temporal profile of large bandwidth lasers and frequency combs can
be manipulated \citep{Weiner_2000}. Other applications in the quantum
optics community, for example, the generation of control signals to
shape magnetic or dipole traps for neutral atoms \citep{Grimm_2000,Folman_2002}
and using SLMs for wave front sensing, adaptive imaging, and holography
also require a large number of individual voltages.

Commercially available devices do not meet the requirements of precise enough voltages combined  
with a fast update rate as needed for  quantum optics experiments.
This article reports on the design, development, and characterization of a 
scalable, fast and multichannel arbitrary waveform generator (MAWG)\citep{Baig_2012} 
that meets the requirements of advanced quantum optics experiments.

\section{Hardware architecture}

\begin{figure}[b]
\begin{centering}
\includegraphics[width=80mm]{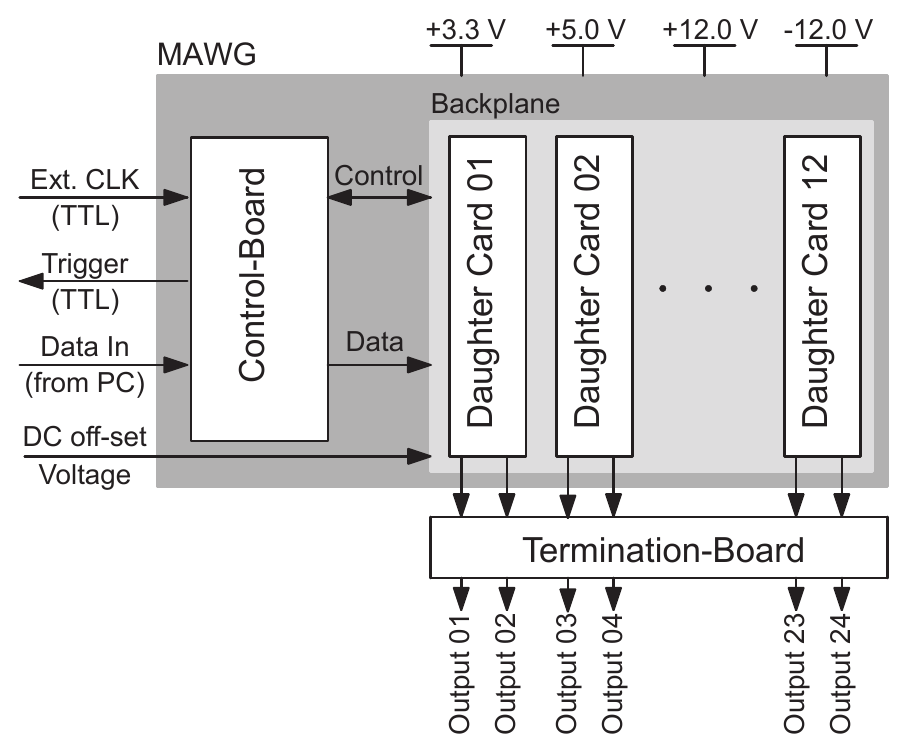} 
\caption{The general architecture of the MAWG showing the inputs, the outputs and flow of data and control signals among main parts of the device.}
\label{fig:BlockDia_Device} 
\par\end{centering}
\end{figure}

A general architecture of the device is shown in Fig.~\ref{fig:BlockDia_Device}.
The MAWG consists of a control-board interfacing with 12 identical
daughter-cards via a backplane. The control-board receives the waveform-data
as well as control instructions from a control computer via a USB\,2.0
interface. A field programmable gate array (FPGA) on the control-board
decodes the control instructions and stores the provided waveform-data
on the memory of a desired daughter-card or retrieves the pre-stored
waveform-data from the memory of the daughter-card according to the
control instruction. The backplane provides distribution of signals
and voltage supplies to all of the daughter-cards. Each daughter-card
has two channels to generate two independent voltages which are individually
terminated and low-pass filtered on a remote termination-board resulting
in the final voltage signals to be used in the desired application. A dedicated synchronous dual-port static RAM (DP-SRAM) is provided
to each of the daughter cards to store waveforms data for every channel
individually up to 128\,k depth. A complex programmable logic device
(CPLD) on every daughter card controls the flow of the data and the clock
to both of the channels locally. The general architecture of the MAWG
with flow of data signals among main parts of the device is shown
in Fig.\,\ref{fig:BlockDia_Device}

\begin{table}[t]
\caption{List of main components used in the MAWG}

\begin{centering}
\begin{tabular}{>{\raggedright}p{2.8cm}>{\raggedright}p{2.8cm}>{\raggedright}p{2.5cm}}
\hline 
\textbf{\scriptsize Component} & \textbf{\scriptsize Company} & \textbf{\scriptsize Part \#}\tabularnewline
\hline
{\scriptsize USB \textmu{}C} & {\scriptsize Cypress Semiconductor} & {\scriptsize CY7C68013A}\tabularnewline
{\scriptsize FPGA} & {\scriptsize Xilinx Inc.} & {\scriptsize XC3S500E}\tabularnewline
{\scriptsize Serial EEPROM} & {\scriptsize Microchip Technology} & {\scriptsize 24LC64}\tabularnewline
{\scriptsize CPROM} & {\scriptsize Xilinx Inc.} & {\scriptsize XCF04S}\tabularnewline
{\scriptsize Crystal} & {\scriptsize Vishay } & {\scriptsize XT49M}\tabularnewline
{\scriptsize Oscillator} & {\scriptsize IQD Frequency Products Limited} & {\scriptsize IQXO-22C}\tabularnewline
{\scriptsize CPLD} & {\scriptsize Lattice Semiconductor } & {\scriptsize ISPMACH4064V}\tabularnewline
{\scriptsize DP-SRAM} & {\scriptsize Integrated Device Technology} & {\scriptsize 70V3599}\tabularnewline
{\scriptsize Voltage Reference} & {\scriptsize Maxim Integrated} & {\scriptsize MAX6161}\tabularnewline
{\scriptsize DAC} & {\scriptsize Maxim Integrated} & {\scriptsize MAX5885}\tabularnewline
{\scriptsize OP-AMP} & {\scriptsize Analog Devices} & {\scriptsize AD8021AR}\tabularnewline
{\scriptsize Digital Isolator A} & {\scriptsize Analog Devices} & {\scriptsize ADUM1100BRZ}\tabularnewline
{\scriptsize Digital Isolator B} & {\scriptsize Analog Devices} & {\scriptsize ADUM1401CRWZ}\tabularnewline
{\scriptsize Octal Bus Transceiver} & {\scriptsize Texas Instruments} & {\scriptsize SN74LVT245B}\tabularnewline
{\scriptsize DC-DC Converter} & {\scriptsize Texas Instruments} & {\scriptsize TPS54610PWP}\tabularnewline
\hline
\end{tabular}
\par\end{centering}

\label{table: components_list}%
\end{table}

The MAWG is optimized to get a good quality signal (in terms of low
reflections, drift and noise) for operations up to an update rate
of 25\,Msps. A user defined internal clock, provided by the FPGA
on the control board, or an external transistor–transistor logic (TTL) signal can be used to operate
the device in an \textit{asynchronous} or \textit{synchronous }mode of operation respectively. In addition, a TTL pulse signal is provided by the MAWG device to trigger the operations of other devices in the experiment. The user-defined voltage waveforms can optionally be shifted by a DC off-set voltage.

The updating of waveforms in the device memory - the \textit{write}
operation - is done sequentially, channel-by-channel with one channel
at once, while the actual generation of voltages - the \textit{read}
operation - is executed simultaneously and synchronously for all of
the 24 channels. Provision of the dedicated memory for each channel
and simultaneous read for all channels correspond to a digital bandwidth
of 9.6 Gbps.

\begin{figure*}[t]
\begin{centering}
\includegraphics[width=165mm]{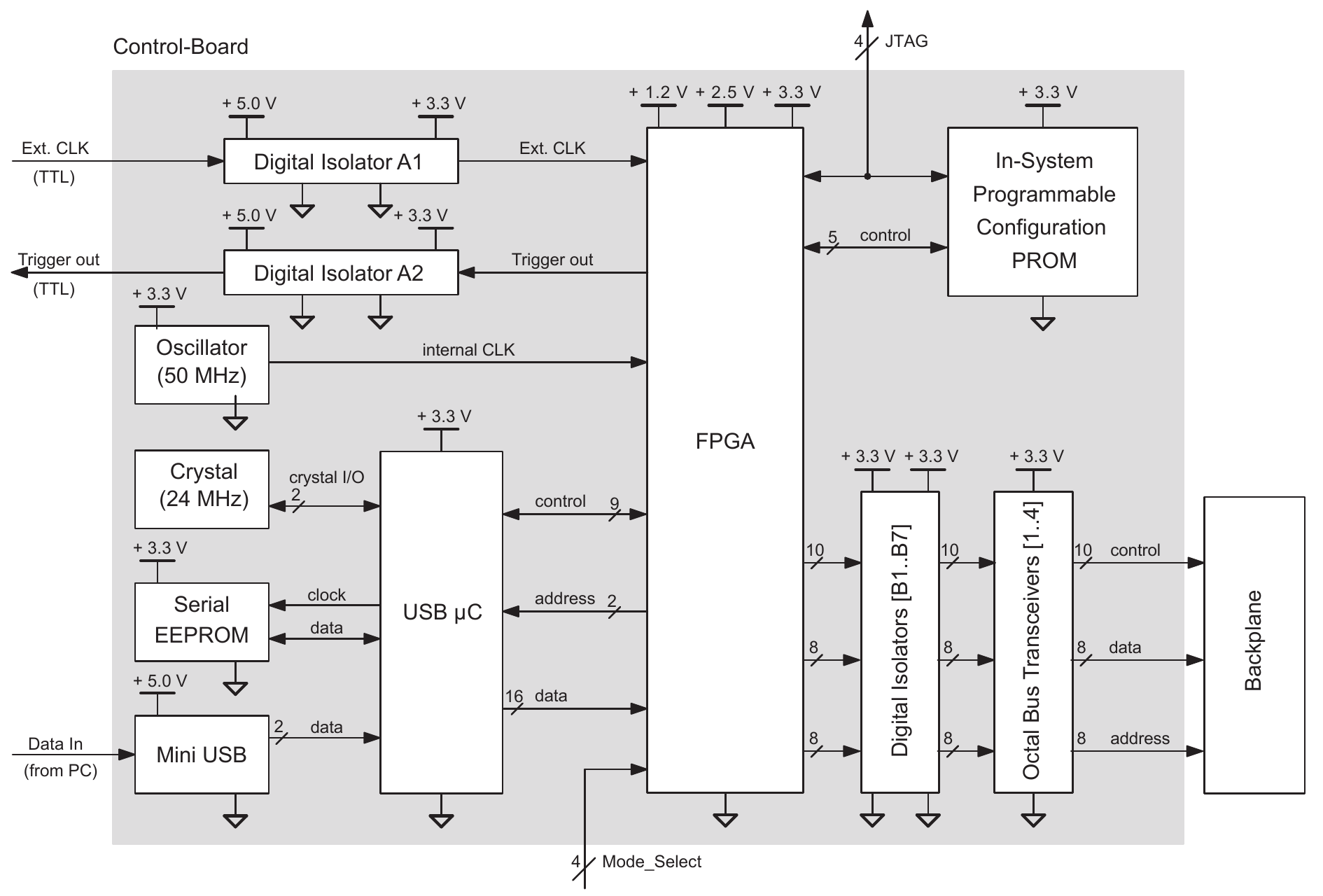} 
\caption{Data and signal flow among the components on the control-board. The
USB \textmu{}C receives the data from the control computer via the
USB and transfers it to the FPGA which delivers the data to the backplane
through the digital isolators and the octal bus transceivers. An optional
TTL input and a TTL output are also provided through the digital isolators.
The CPROM and the EEPROM are provided to auto-configure the FPGA and
the USB \textmu{}C respectively.}
\label{fig:BlockDia_ControlBoard} 
\par\end{centering}
\end{figure*}

A list of the main integrated circuit components (ICs) used for the
MAWG device is provided in Table\,\ref{table: components_list}. \footnote{All devices and components 
listed are specified for traceability purpose only and are not mentioned 
for advertising them, nor do we claim superior performance.}
All components were chosen to satisfy various assessment criteria,
importantly, high dynamic range of speed and amplitude, supportive
contribution to low-noise and environmental stable performance, as
well as possibly general simplicity and high reliability at the same
time.

\subsection{The control board}

The control-board holds both fixed- and firmware-type circuitry, which
are primarily needed to receive user-defined waveform and/or control
instructions sent from the control computer, and furthermore to process
and to deliver them to the daughter-cards through the backplane in
a controlled way.

The data is received from the control computer by the USB micro-controller
(\textmu{}C) via the USB interface and delivered to the FPGA in a
2-bytes word format. The FPGA analyses the data and delivers a single-byte-oriented
output either directly or using its internal block-RAM to the backplane
in case of the instructions or the waveform-data respectively. All
outputs of the control-board are galvanic-isolated followed by a signal-current
enhancement provided by octal bus transceivers. A configuration programmable
read-only memory (CPROM) and a serial electrically erasable programmable
read-only memory (EEPROM) are used to auto-configure the FPGA and
the USB \textmu{}C after power-up respectively. The components of
the control-board and the data flow among them are detailed in 
Fig.\,\ref{fig:BlockDia_ControlBoard}.

All voltage supplies required by the components on the control-board
(1.2\,V, 2.5\,V, 3.3\,V and 5.0\,V) are down-converted on-board
from a single +5.0\,V supply using DC-DC converters. A galvanic isolated
TTL compatible input and output is provided to synchronize operation
of the device with rest of the devices in the experiment. The TTL
input can be used either as a trigger for internal clock or to apply
a signal as an external clock for the synchronized use of the MAWG.

A 24\,MHz crystal is used by the USB \textmu{}C to generate internally
480\,MHz and 48\,MHz clocks needed for USB data transmission. Furthermore,
a 50\,MHz on-board oscillator is used by the FPGA to generate clock
signals by means of internal frequency divisions, which are needed for the 
\textit{write} and the \textit{read} operations of the MAWG

\subsection{The backplane with the daughter-cards}

\begin{figure*}[ht]
\begin{centering}
\includegraphics[width=170mm]{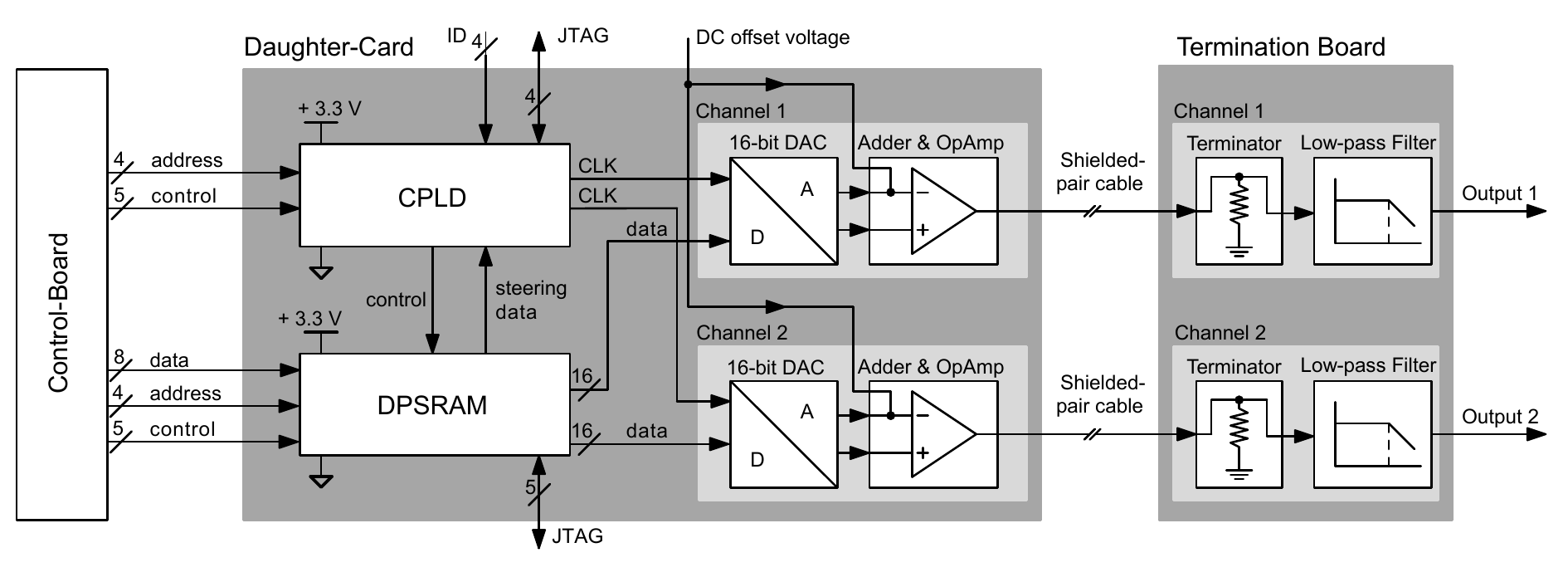} 
\caption{Block diagram of the daughter-card circuitry. A flow of the digital 
data and the control signals from the FPGA on the control-board to the CPLD 
and the DP-SRAM on the daughter-card is laid-out in detail followed by block 
diagrams of the the channels, the termination resistors and the low-pass filters.}
\label{fig:BlockDia_DaughterCard} 
\par\end{centering}
\end{figure*}

\begin{figure*}[ht]
\begin{centering}
\includegraphics[width=170mm]{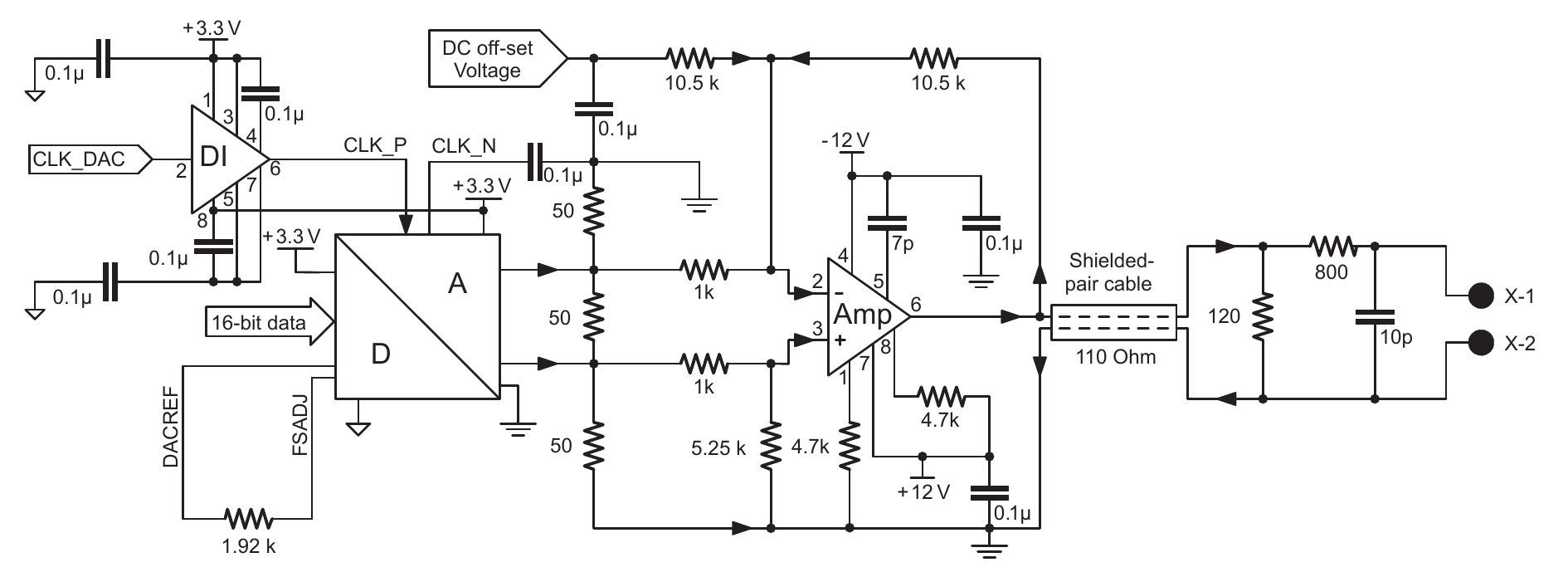} 
\caption{A detailed circuit diagram for a channel on the daughter-card. 
The digital signals from the DP-SRAM are converted to the analog signals
in the DAC followed by an optional addition of the DC-offset at the
differential OP-AMP with an amplification of 10. Downstream the output
is terminated and low-pass filtered. The digital isolator provides
the noise-free clock signal to the DAC.}
\label{fig:Schematic_Channel} 
\par\end{centering}
\end{figure*}

The backplane holds 12 identical daughter-cards such that each daughter-card
shares a common data-, control- and address-bus. The waveform-data
received from the control-board is delivered to a daughter-card via
the backplane during the \textit{write} operation of a channel on the
daughter card. All bus lines are impedance-matched and resistor-terminated
on the backplane. Three supply voltages +3.3\,V and \textpm{}\,12\,V
(optionally \textpm{}\,15\,V) generated by a dedicated external
power supply are distributed to each daughter-card trough the backplane.

All power lines are protected by Zener diodes at the entrance of the
backplane.

The daughter-card contains the circuitry to generate two independent
voltages. The key components used on the daughter-card along with
a flow of data and control signals are shown in Fig.\,\ref{fig:BlockDia_DaughterCar}.
The CPLD controls signals received by the card and generates clock
pulses for the digital-to-analog converter (DAC) along with the control
signals to the DP-SRAM on the same daughter-card. The control signals
manage storage of the data to a particular address of the DP-SRAM
when a \textit{write} operation of the MAWG is executed or deliver the
stored data from a particular address of the DP-SRAM to the desired
DAC when a \textit{read} operation of the MAWG is performed. The CPLD
and the DP-SRAM serve for both channels of the respective daughter-card.
Each 16-bit DAC converts respective digital code into a pair of analog
signals and delivers it to the operational amplifier (OP-AMP). The
output voltages provided by a daughter-card are coupled to a shielded-pair
cable with 110\,$\Omega$ characteristic impedance. The key components
of the daughter-card are listed in Table\,\ref{table: components_list}.

A channel on the daughter-card can be operated in two modes of operation,
\textit{stopped-clock} and \textit{continued-clock}. In the \textit{stopped-clock}
mode the DAC of the channel holds last value of the voltage sequence
being read from the DP-SRAM without getting further clock signal from
the CPLD whereas in the \textit{continued-clock} mode the DAC gets a
continuous clock and refreshes the last value of the voltage sequence
being read from the memory with the user-defined clock update rate.

\subsection{A channel on the daughter-card\label{sub:A-Channel}}

A single channel on a daughter-card consists of a complementary current-type-output
16-bit DAC, followed by a low-noise, high-speed OP-AMP in differential
configuration with a gain equal to 10. The DAC converts the user defined
voltage code, pre-stored in the DP-SRAM using the \textit{write} operation,
into two corresponding analog current signals with a 16\,bit resolution.
These current signals are then converted into two voltage drops across
50\,$\Omega$ resistors, referenced to the analog ground (AGND).
The voltage drops are finally subtracted and amplified through the
OP-AMP circuit. An optional addition of a DC-offset voltage of up
to \textpm{}\,7\,V\textsubscript{pp} is also provided which is
common to all channels and can be used to shift the reference level
of the final output provided that the sum of amplitudes of the actual
and offset voltages does not exceed the output amplitude limits of
the channel (see Fig.\,\ref{fig:Schematic_Channel}).

The DAC is configured to use the full scale output current amplitude (\textpm{}\,2\,mA) and to use an internal band-gap reference voltage (1.2\,V) provided by the DAC itself. The output of the DAC has a latency of 3.5 clock cycles.

\subsection{The memory management on the daughter-card}

The DP-SRAM has a total capacity of 128\,k words with a word-width
of 36\,bits. The DP-SRAM is organized by dividing the full word of
the memory into two sub-words of 18\,bits each, such that each word
serves to hold one data code of a channel on the respective daughter-card
by using 16\,bits, whereas the remaining 2\,bits are reserved for
optional steering data. The full depth of the DP-SRAM is divided into
16 logical segments of 8\,k words each, which in practice reduces
the width of address bus to access the memory from 16 to 4 bits and
hence results in a reduced digital noise on the daughter-card at the
expense of discrete access to the memory only at start-addresses of
the 16 segments \citep{Baig_2010_DPG}.

\begin{figure}[th]
\begin{centering}
\includegraphics[width=80mm]{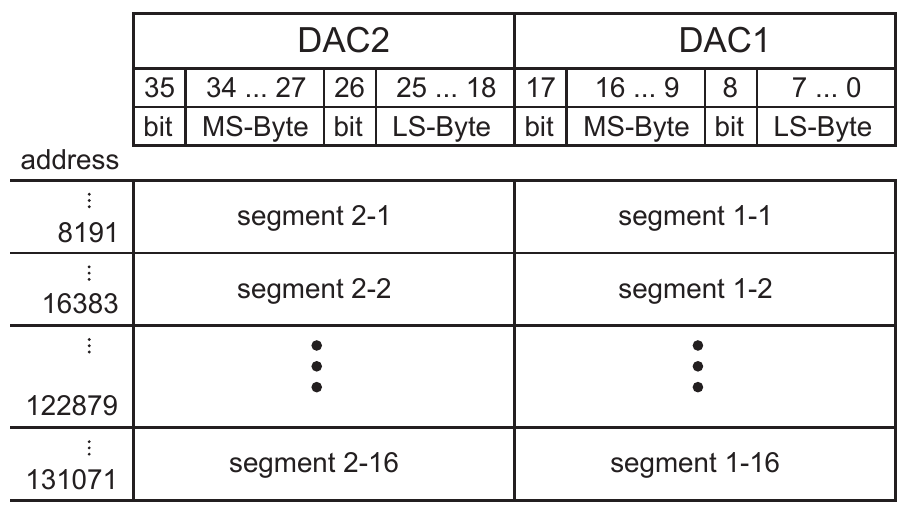} 
\caption{Memory management scheme in the DP-SRAM of the daughter-card. Each
word of the memory is divided into two sub-words serving for both
channels independently while depth of the memory is divided into 16
logical segments. A bit-assignment for one 36\,bits memory word is
also shown with indication of four reserved bits to be used for steering
the data. Each memory segment holds 18432 bytes.}
\label{fig:SRAM_Model} 
\par\end{centering}
\end{figure}

Each segment of the DP-SRAM can be used to store an individual voltage
sequence of up to 8\,k words. A voltage sequences larger than the
size of the segment can be stored on multiple adjacent segments until
the end of the memory. Once a memory segment is accessed at the beginning
of the \textit{write} or the \textit{read} operation, the address-control
is passed to the DP-SRAM for further increment of the address, using
its internal counter, which can be continued across the segments until
the end of the memory. Fig.\,\ref{fig:SRAM_Model} shows details
about the memory address management and the bit-assignment of one
36\,bits memory word. The waveforms stored on different segments
can be retrieved with a switching latency of 4 clock cycles from one
segment to another (e.g., 160\,ns for 25\,MHz update rate). Writing
up the data for full depth of the DP-SRAM of a channel needs around
32\,ms.

\subsection{Printed circuit boards}

The printed circuit boards (PCB) for the control-board and the backplane
are designed with 4-layers, while the daughter-cards are designed
with 6-layers architecture. Routing of the signals is managed on top
and bottom layers, whereas the inner layers are reserved only for
power and ground planes with solid and large copper areas, which ensure
a good high-speed performance due to provision of shortest return
current path for signals, low thermal resistance for heat dissipation,
reduced parasitic inductance and up to some extent electromagnetic
shielding\citep{Ott_2001,Ardizzoni_2005}.

The partitioning of the digital and the analog ground- and power-planes
keeps them separated to make it sure that rich in noise digital circuitry
does not affect the performance of the analog part 
\citep{Johnson_1993, Ott_2001, Mancini_2002_TexasInstr., Rizvi_2005, Hu_2005}.
On the other hand, the digital and the analog grounds are connected
to a single common point with a high impedance resistor to avoid dipole
and loop antenna effects due to a difference of potentials and large
return path respectively \citep{Ott_2001,Rizvi_2005,Ardizzoni_2005}.

Routing of the digital and analog signals is done with extensive care,
such that not even a single digital signal trace is crossing the analog
ground or power plane and vice versa. In addition to the careful routing,
the clock signal to the DAC is routed through a galvanic isolator
(that bridges analog and digital grounds and power planes) to maintain
the current-loop separation strictly. In addition to the general mixed-signal
design recommendations, the particular PCB design recommendations for the DAC 
and OP-AMP are also followed to get optimum performance. In particular, the OP-AMP 
is ensured to have the shortest feedback connection and ground-free area under the
signal input traces to avoid ringing and overshoot, as well as to
reduce a stray input capacitance\citep{Ardizzoni_2005}.

\subsection{Power distribution network}

The distribution of electrical power for all IC-components is provided
to the daughter-card PCB via dedicated inner planes with large surface
and hence low impedance. A hierarchical network of decoupling capacitors
is implemented to ensure availability of large amounts of current slowly
by bulk capacitors as well as less amounts of current rapidly by small
capacitors \citep{Mancini_2002_TexasInstr., Knighten_2005} required by 
the components on the control-board and the daughter-card PCBs.\textbf{ }
A low-pass filter with RF inductance and ad two capacitances (of 10\,\textmu{}F 
and 1\,\textmu{}F) in parallel combination are used at each power supply injection 
into the daughter-card PCB. A decoupling capacitor of 100\,nF
is placed as close as possible to each power pin of all ICs which
decouples the high frequency noise from the analog ICs and work as
a charge reservoir for digital ICs to provide the frequently required
charge during switching of digital components to avoid the corruption
of logic levels\cite {Mancini_2002_TexasInstr.}.

\begin{figure}[t]
\begin{centering}
\includegraphics[width=80mm]{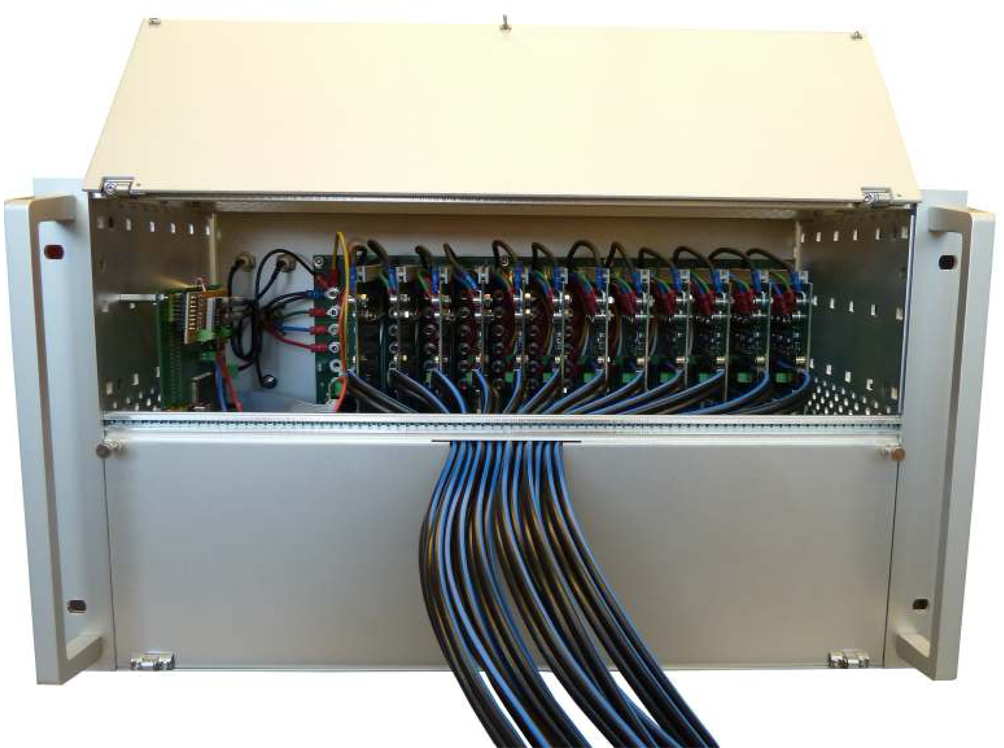} 
\caption{The MAWG device showing the control-board (on left) and the backplane
housing 12 daughter cards on it. }
\label{fig:MAWG_device} 
\par\end{centering}
\end{figure}

\begin{figure}[thb]
\begin{centering}
\includegraphics[width=80mm]{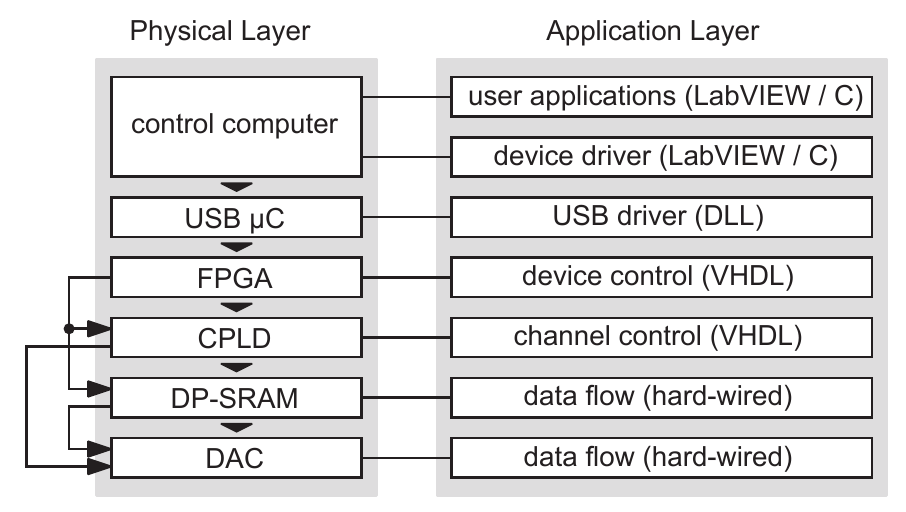} 
\caption{Software architecture model of the MAWG showing the physical and 
the application layers along with their inter-connections. The arrows from 
top to bottom show data flow whereas arrows on left side show flow of control among components of the physical layer.}
\label{fig:SW_Model} 
\par\end{centering}
\end{figure}

\begin{figure}[thb]
\begin{centering}
\includegraphics[width=80mm]{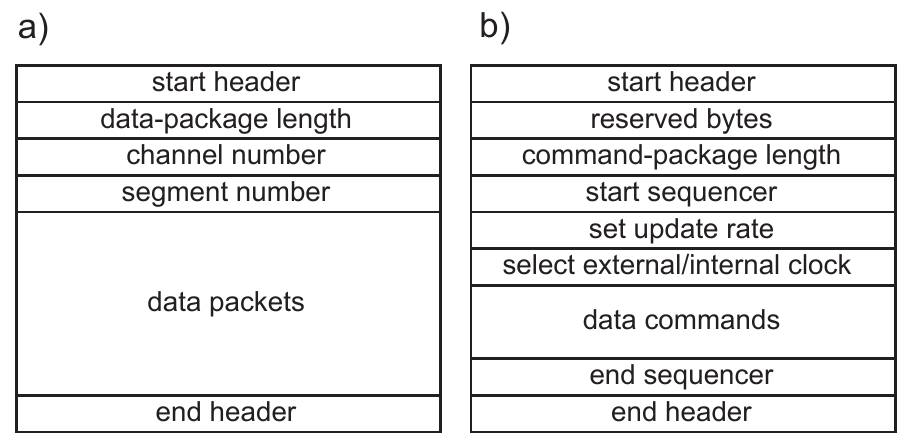}
\caption{Protocols defined to send the data-package (a) and the command-package
(b) from the control computer to the MAWG.}
\label{fig:CommunicationProtocols} 
\par\end{centering}
\end{figure}

\begin{table*}[bth]
\caption{Exemplary sequence of  commands to get the waveform data pre-loaded in
the DP-SRAM. The code for the commands consists of command identifier
(number on left) and parameter for action (rest of the number/numbers).}
\begin{centering}
\begin{tabular}{>{\centering}m{1cm}ll>{\raggedright}m{12cm}}
\hline 
Implicit Index  & Command  & Code  & Explanation\tabularnewline
\hline 
\hline 
 & StartSequencer  & 32 : 0  & Start execution of the command sequencer.\tabularnewline
 & SetUpdateRate  & 34 : 2  & Set the update frequency equal to 25 MHz (50\,MHz divided by 2).\tabularnewline
 & ExtCLK\_Disable  & 35 : 0  & Enable internal clock for read operation of the MAWG.\tabularnewline
0  & WaitForPulse  & 19 : 0  & Wait for external trigger to start the sequencer.\tabularnewline
1  & StartSegment  & 02 : 7250  & Output 7250 words starting from beginning of the segment 2 of the
DP-SRAM.\tabularnewline
2  & Pause  & 16 : 255  & Pause the execution for 255 clock cycles. \tabularnewline
3  & StartSegment  & 05 : 10500  & Output 10500 words starting from beginning of the segment 5 of the
DP-SRAM.\tabularnewline
4  & Repeat  & 17 : 0 : 11  & Repeat 11 times the execution of preceding commands starting from
index 0.\tabularnewline
5  & SendPulse  & 18 : 170  & Issue an output pulse for 170 clock cycles for external synchronization.\tabularnewline
 & EndSequencer  & 41 : 0  & End execution of the command sequencer.\tabularnewline
\hline 
\end{tabular}
\label{table:datacmdExample} 
\par\end{centering}
\end{table*}

Towards the superior goal of reducing the noise voltage in the system,
we took several layout related practical measures, among them we satisfied
the following key recommendations for mixed-signal PCB design 
\citep{Smith_1999, Kester_2000, Mancini_2002_TexasInstr., Fan_2002_IEEE, Archambeault_2007, Archambeault_2008, Archambeault_2009}:
(1) Selecting ceramic decoupling capacitors with small SMD package and low equivalent series inductance. 
(2) Placing decoupling capacitors as close as possible to the power pins of the ICs (\textasciitilde{}4\,mm).
(3) Maximizing power-to-ground inter-plane capacitance by minimizing the distance between the power and ground planes (\textasciitilde{}0.4\,mm).
(4) Eliminating unwanted capacitance between digital power- and analog ground-planes (and vice versa) by avoiding their geometrical overlap.

Since low inductance path from decoupling capacitors to ground plane
enhances effectiveness of noise bypassing, its implementation was
pursued consequently with high priority during the PCB layout. Based
on calculations given in \citep{Archambeault_2009}, the above design
considerations would result in an expected inductance of about 2\,nH
per single decoupling capacitor and a power pin of the IC.

Fig.\,\ref{fig:MAWG_device} shows the fully assembled MAWG with the
control-board (on left side in the box) and 12 daughter cards mounted
on the backplane.

\section{Software architecture}

The software needed to run the MAWG was developed in VHDL, C and LabVIEW.
The firmware development for the FPGA and the CPLD was done in VHDL
and both of the devices are burned with the tested programs using
software interfaces provided by the respective vendors. A high-level
user-interface to control the MAWG parameters at the run-time is designed
in the C as well as in the LabVIEW programming languages. Both applications
provide graphical user interface (GUI) for easy access to the MAWG
device. Fig.\,\ref{fig:SW_Model} shows a model-based architecture
of the developed software along with the flow of information from
one physical- and application-layer component to the other one.

The user communicates with the MAWG device by means of sending data-packages
for writing into the DP-SRAM and subsequently command-packages for
execution in a command sequencer (CS), designed and implemented inside
the FPGA. Both communication packages follow a pre-defined data and
command bit-formats.

The data-package protocol (Fig.\,\ref{fig:CommunicationProtocols}a)
is used to send the data to be written on the DP-SRAM of a daughter-card.
It starts with a start-header followed by information of length of
the data-package, the channel number and the segment number to be
written. Then comes the data in form of packets of 9\,B each and
at last the end-header to stop the process of writing the DP-SRAM.

\begin{figure*}[th]
\begin{centering}
\includegraphics[width=161mm]{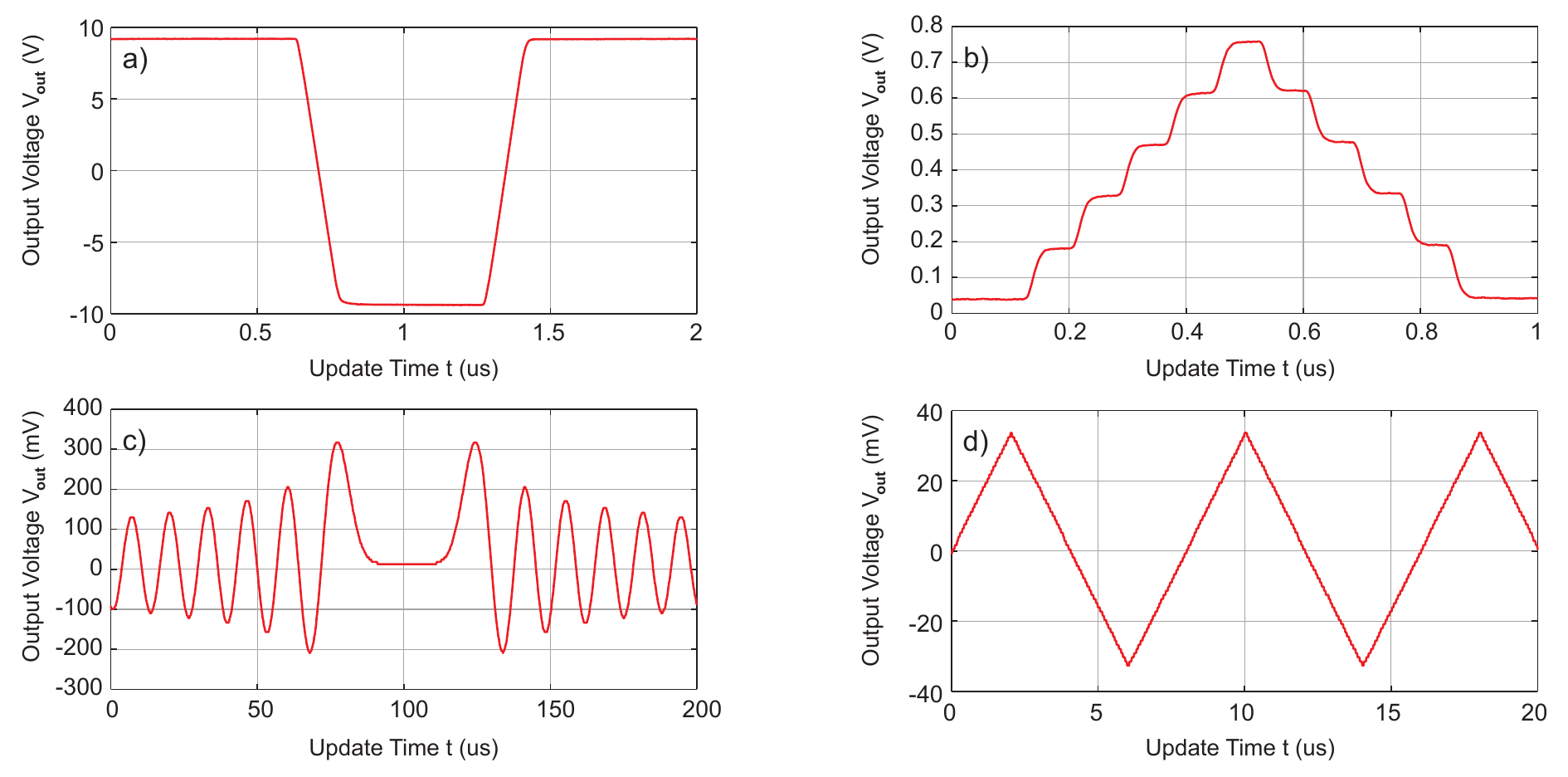} 
\caption{Versatile waveforms generated by the MAWG : a) an abrupt changed voltage
with a large slew rate, and no visible ringing, b) a sequence of small
fast voltages steps, c) generation of arbitrary waveforms (here
a Bessel function) and d) a triangular waveform with a good linearity
of output}
\label{fig:waveforms}
\par\end{centering}
\end{figure*}

Similarly the command-package protocol (Fig.\,\ref{fig:CommunicationProtocols}b)
consists of instructions to read a pre-stored data from the DP-SRAM
and make it available for the DAC. It starts with a start header followed
by the length of the command package and the instruction to start
the sequencer which encapsulates a list of commands followed by the
instruction to end the sequencer and the header. The list of commands
consist of configuration commands followed by a combination of five
data-handling commands. The configuration commands, \textit{SetUpdateRate}
and \textit{ ExternalClock,} configure the clock update
rate for the internal clock or enable/disable the external clock for
the \textit{read} operation of the device respectively. The data-handling
commands, \textit{StartSegment, Pause, Repeat, WaitForPulse} and \textit{SendPulse}, manipulate the data according to user defined combination
of the these commands. All data-handling commands have an implicit
index starting with 0. An exemplary command sequencer\footnote{Further details about 
the commands to communicate with device can be provided by the authors on request.} is given in
Table\,\ref{table:datacmdExample}.

\section{Performance}
\subsection{Generation of arbitrary waveforms}

A few exemplary chosen demo-waveform signals generated by the MAWG
are displayed in Fig.\,\ref{fig:waveforms} to demonstrate capability
of the MAWG. Fig.\,\ref{fig:waveforms}a presents an abrupt change
of voltage from the minimum to the maximum accessible value which
shows the amplitude span, the large slew rate, and no visible ringing.
Fig.\,\ref{fig:waveforms}b presents a sequence of small fast voltages
steps with 80\,ns dwell time demonstrating the large bandwidth. 
Fig.\,\ref{fig:waveforms}c shows the generation of an arbitrary waveforms 
(here a Bessel function). Fig.\,\ref{fig:waveforms}d presents a 
triangular waveform showing a good linearity of the output.

\subsection{Output voltage range}

\begin{figure*}[th]
\begin{centering}
\includegraphics[width=160mm]{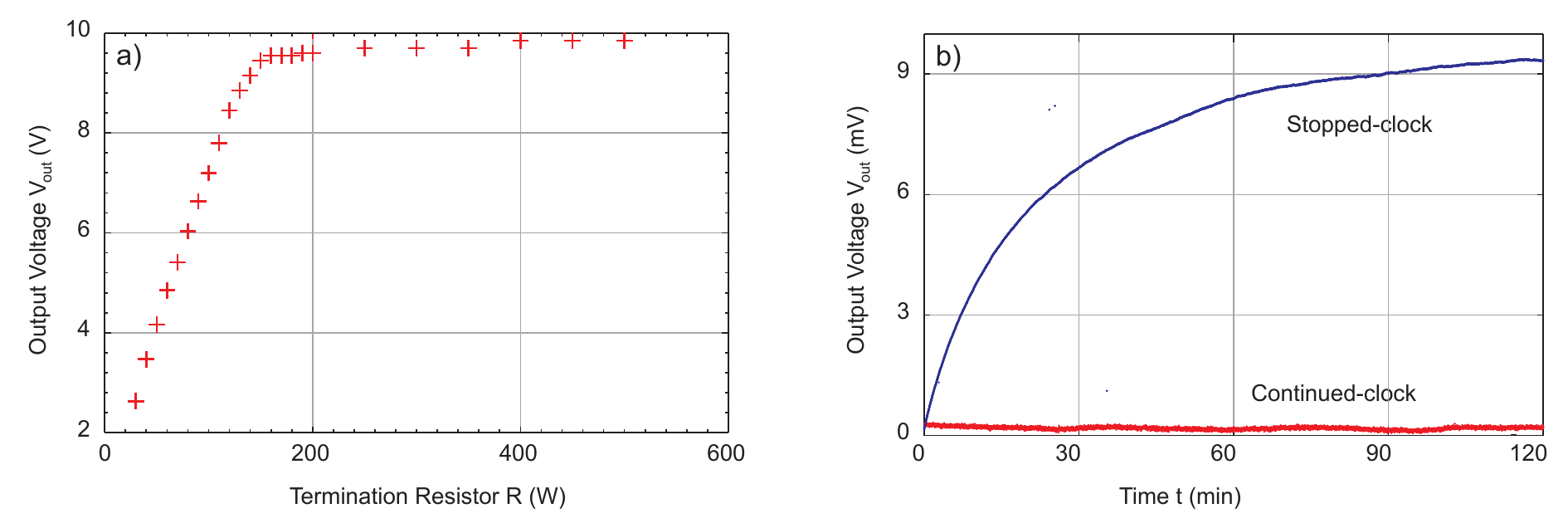} 
\caption{(a) The output voltage of a channel of the MAWG as a function of termination
resistor, showing the saturation of the output voltage after a certain
resistor value. (b) Output voltage drift of a channel of the MAWG
for the \textit{stopped-clock} and the \textit{continued-clock} modes
of operation. The \textit{continued-clock} mode is used with 1\,MHz
update rate which is the minimum required update rate for the DAC.}
\label{fig:Termination_Drift} 
\par\end{centering}
\end{figure*}

The low-noise OP-AMP (AD8021) has a linear output current limit of 70\,mA 
which determines the maximum output voltage amplitude depending on the value 
of the termination resistor. A resistor value that needs the amount of 
current larger than the specified current limit of the OP-AMP will cause 
saturation of the output voltage. On the other hand, for higher bandwidth 
output a proper impedance matching between the shielded-pair transmission-line 
(with characteristic impedance of 110\,$\Omega$) and the termination resistor 
value on the end of the line is obligatory to minimize the output-signal distortion
due to signal reflections. As depicted in Fig.\,\ref{fig:Termination_Drift}a
the impedance matching scheme with 120\,$\Omega$ termination resistor
provides a maximum signal amplitude range of \textpm{}8.5\,V.\textbf{
}Though, for low-speed signals of less than 1\,Msps\textbf{, }the
termination resistor can be increased to 333\,$\Omega$, resulting
in a less demand for the driving current and consequently providing
higher maximum signal amplitude of nearly \textpm{}10\,V practically
without compromising the signal waveform quality. The measured output
impedance of the MAWG channel, determined from Fig.\,\ref{fig:Termination_Drift}a
at half of the maximum signal amplitude, amounts to 65\,$\Omega$.

Another OP-AMP (THS4631D) is tested to provide higher
current and hence higher output voltage resulting in maximum output
voltages of \textpm{}13\,V and \textpm{}11\,V for the low- and high-speed
signals respectively and an output impedance of 45\,$\Omega$ and
a thermal dependence of around 200\,\textmu{}V/\textdegree{}C. The
OP-AMP is equipped with \textpm{}\,15\,V supply and configured with
a gain of 13.

The characterization of the device and results presented are based
on the OP-AMP (AD8021) with a termination resistor of 333\,$\Omega$.

\subsection{Output voltage precision}

The MAWG device has a linear output signal amplitude behaviour over
full range of the operation. A discrepancy between the expected and
obtained voltages is found to be below \textpm{}\,4\,mV with a standard
deviation of 0.81\,mV (possibly limited by the measurement device).
The discrepancies vary randomly among different channels due to manufacturing
imperfections of the components and can be compensated by a software
calibration of each channel. The main contribution to the output voltage
discrepancy is the offset of the output voltage, which results from
an intrinsic OP-AMP input-offset-voltage of typically \textpm{}\,0.4\,mV
which results in \textpm{}\,4\,mV after amplification.

\subsection{Output voltage long-term stability}

The MAWG device is characterized by dynamic update rates from 25\,Msps
down to DC-signal generation. Long-term stability of DC output signals
is of particular interest for applications in quantum information
processing. The static operation of the MAWG channel is determined
mainly by the performance of our DAC component, which is specified
by the vendor for output signal update rates from 1\,Msps to 200\,Msps.

In the \textit{stopped-clock} mode of operation the updating of the
DAC is suspended during static output voltages, resulting in an idle
state of digital switching. This mode reduces the clock noise on the
output signal at the expense of a voltage drift of up to several mV
for long term operation with some variations among different channels.
On the other hand, in the \textit{continued-clock} mode of operation,
the DAC is updated continuously with a user-defined clock and hence
gives an output voltage with higher stability, however with a tradeoff
on noise.

An output signal fluctuating around the set voltage with a maximum
deviation less than \textpm{}\,150\,\textmu{}V over a period of
2\,h and long term drift with a slope below 10\,\textmu{}V/min is
observed when the clock is operated with the minimum specified clock
update rate of 1\,MHz (Fig.\,\ref{fig:Termination_Drift}b). A combined
operation of both modes can be used to get optimum performance. For
example, in a particular experiment, the device can be operated in
the \textit{stopped-clock} mode to get the low-noise output during the
measurement and then can be switched to the \textit{continued-clock}
mode to get the low-drift output during waiting time between two consecutive
measurements. 

The overall thermal dependence of the MAWG output voltage is around
70\,\textmu{}V/\textdegree{}C and is measured by a gradual increase
of temperature by up to 14\,\textdegree{}C starting from the ambient
air temperature.

\subsection{Output voltage noise level}

\begin{figure*}[th]
\begin{centering}
\includegraphics[width=170mm]{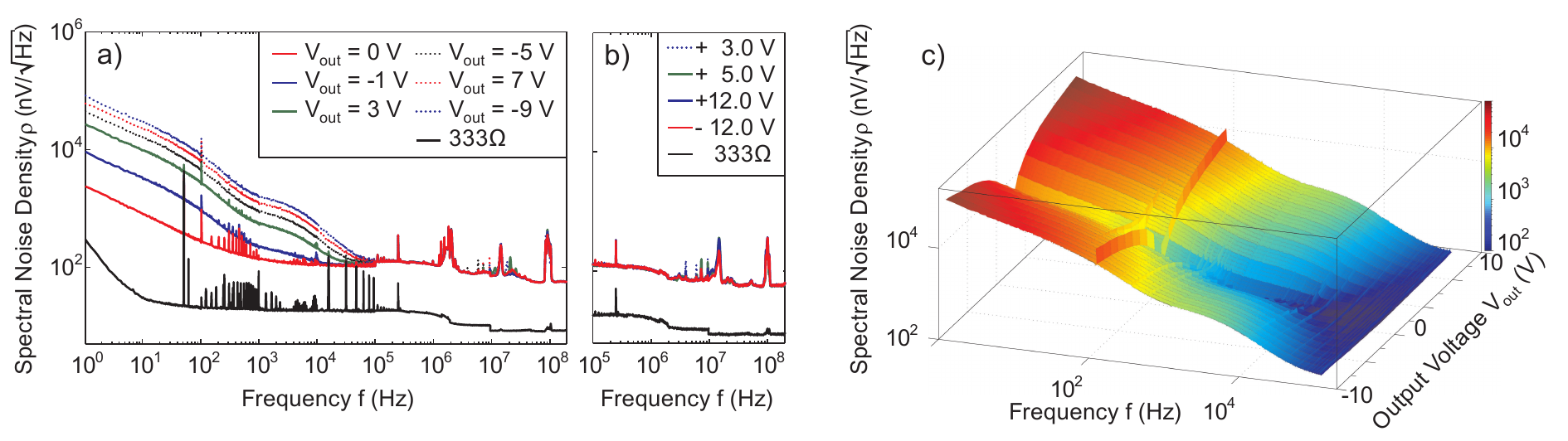} 
\caption{a) The spectral noise density measured for the output of one channel
of the MAWG, in the \textit{stopped-clock} mode of operation, with the
termination resistor but without the low-pass filter. The spectra
up to 100\,kHz and higher frequencies are taken by two devices with
an average of 1000 and 10000 respectively. The black spectrum represents
the measurement noise floor obtained by taking the spectra across a naked resistor.
b) The SND measured at the input of the MAWG (i.e., contributions 
from power supply and pick up from externally generated radio frequencies) explaining the noise peaks at higher frequencies present in the spectral noise density profile at the output of the MAWG. c) The spectral noise density measured as a function of frequency and output voltages for the frequency range below 100\,kHz.}
\label{fig:SND_2D_3D} 
\par\end{centering}
\end{figure*}

The analysis of the MAWG output voltage noise is based on the spectral
noise density (SND) $\rho$ of the device measured with the \textit{stopped-clock}
mode of operation. The SND of the MAWG is determined for the full range
of its output voltages by taking several voltage spectra for different
DC output voltages, whereas a wider spectral range from 1\,Hz to
200\,MHz was covered by using two different spectrum analysers
\footnote{Advantest (R9211B) and Rohde \& Schwarz (FSP 30)}
and later concatenating the resulting spectra for each of the cases
(Fig.\,\ref{fig:SND_2D_3D}a). The SND spectra are calculated using
the relation

\[
\rho=\sqrt{\frac{\sigma_{\mathrm{v}}^{2}}{RBW}}
\]

where $\sigma_{\mathrm{v}}$ is the noise voltage and RBW is the resolution
bandwidth of the spectrum analyser for a particular frequency scan.

\begin{figure*}[th]
\begin{centering}
\includegraphics[width=170mm]{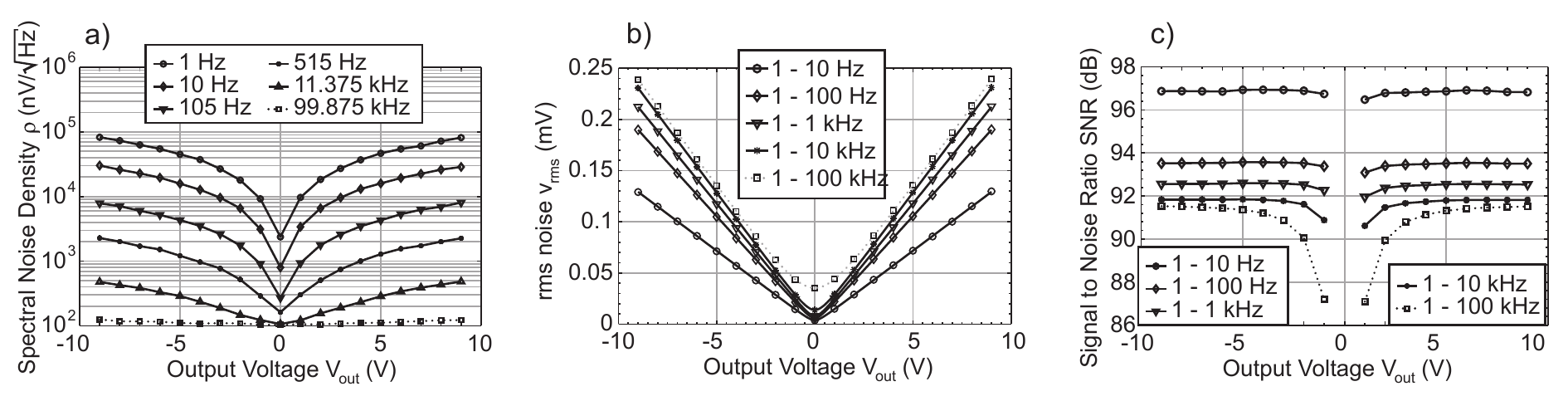} 
\caption{a) The spectral noise density as a function of the output voltages
measured at different frequencies (avoiding the frequencies corresponding
to the 50\,Hz harmonics. b) The rms noise as a function of output
voltages measured at some random frequency spans. c) The signal to
noise ratio as a function of output voltages measured at the frequency
spans. The plots with dotted line and square markers in (a) and (b)
\& (c) represents the highest frequency and the highest frequency
span respectively.}
\label{fig:SND_RMS_SNR} 
\par\end{centering}
\end{figure*}

The SND spectra of the output of a channel of the MAWG, in the \textit{stopped-clock} mode 
of operation, as a function of the frequency are shown in Fig.\,\ref{fig:SND_2D_3D}a for 
a couple of positive and negative output voltages. The spectrum labelled as ``333\,$\Omega$'' 
is taken using the bare load-resistor and represents noise contributions from the measurement.

The spectra show discrete contributions mainly from 50\,Hz and its
harmonics in low frequencies. Some additional noise components are
visible at frequencies higher than 1\,MHz which are identified as
a combined effect of power supply output and pick-up of externally
generated radio frequencies. The smooth noise floor of the output
voltage has a direct dependence on absolute output voltage up to a
frequency of 100\,kHz whereas for rest of the observed spectrum this
dependence is negligible. A measurement for the MAWG input noise density
including power supply output and the 2\,m long, multi-pole, interconnect
cable is plotted in Fig.\,\ref{fig:SND_2D_3D}b which explains the
peaks present in the spectra at higher frequencies and hence can be
avoided by using a storage power supply and/or implementing a proper
low-pass filter at output of the MAWG. In the following analysis,
contributions appearing above 100\,kHz are therefore neglected.

The increasing noise floor with increasing output voltage (Fig\,\ref{fig:SND_2D_3D}c)
is caused by an asymmetry between two complementary output currents
of the DAC (see section \ref{sub:A-Channel}) which are symmetric
for 0\,V output and become asymmetric otherwise. The asymmetry increases
proportionally to the absolute value of the output voltage. A noise cancellation
between complementary outputs occurs in the differential OP-AMP, with
a maximum noise reduction for the symmetric case and a lower otherwise.
On the other hand, the noise floor for each of the output voltage
settings decreases strongly with rising frequency. In Fig.\,\ref{fig:SND_RMS_SNR}a
the SND is shown as a function of channel output voltage at a couple
of frequencies (other than the frequencies corresponding to the 50\,Hz
harmonics).

The root-mean-square noise $\mathrm{v_{rms}}$ of a channel of the
MAWG is calculated using the relationship

\[
\mathrm{v_{rms}}=\sqrt{{\displaystyle \sum_{i=1}^{N-1}}\left\langle \rho_{i,i+1}^{2}\right\rangle \triangle f_{i,i+1}}
\]

where $\triangle f$ is the difference of two consecutive frequencies
obtained in the SND measurements.

The output voltage noise $\mathrm{v_{rms}}$ as a function of the
output voltage ($\mathrm{V_{out}}$) for different frequency spans
is plotted in Fig.\,\ref{fig:SND_RMS_SNR}b which shows a linearly
increased $\mathrm{v_{rms}}$ with increase in output voltage amplitude
of the MAWG and reaches a maximum level of nearly 250\,\textmu{}V
at \textpm{}9\,V for the frequency span of 100\,kHz. Similarly an
increased frequency span results in a corresponding increase in the
root mean square noise which remains consistent for all different
output voltages.

The signal to noise ratio (SNR) of the MAWG output is calculated for
different output voltages using the relationship

\[
\mathrm{SNR}=10\mathrm{\cdot log_{10}}\left(\frac{\mathrm{V_{out}}}{\mathrm{\mathrm{v}_{rms}}}\right)^{2}
\]

The SNR as a function of output voltage $\mathrm{V_{out}}$ is plotted
in Fig.\,\ref{fig:SND_RMS_SNR}c for the same frequency spans. The
dependence of the SNR on the output voltage is more pronounced for
higher bandwidth whereas for constant bandwidth the dependence is
strongest on smaller voltages. The bandwidth of 100\,kHz has a SNR
greater than 90\,dB for the output voltages larger than 2\,V.

\section{Conclusions}

The MAWG is a general purpose DAC-based system designed to provide
multichannel voltage signals characterized by the high dynamic range of
both signal amplitude and signal update rate, as well as by a low
signal noise and a very good long-term signal stability.

The MAWG  is characterized by a large range of signal amplitudes of up to \textpm{}9\,V, a wide dynamic range of update rates from 25\,Msps down to DC signal generation,
as well as a low signal-to-noise ratio of +90\,dB and long-time stable
output signal amplitude within \textpm{}0.3\,mV. The user-control of the device 
is provided by means of high-level PC-software applications,
as well as device firmware, for storing pre-defined voltage sequences
and for generating output signals in real-time. The overall modular
concept of the hardware and embedded software allows for a broad flexibility
in various operation schemes and particularly in the performance tuning.

Although the detailed specifications of the device are defined in response to
technical challenges of experimental quantum information processing,
its practical use for other applications in the fields of fundamental
and applied research is easily conceivable, especially in those of
similar needs for outstanding instrumentation performance.

\section{Acknowledgment}

We acknowledge funding from the European Community's Seventh
Framework Programme (FP7/2007-2013) under Grant Agreement No. 270843
(iQIT) and No. 249958 (PICC) and European Metrology Research Programme (EMRP) 
which is jointly funded by the EMRP participating countries 
within EURAMET and the European Union.

The authors would like to thank Mr. Julius Krieg for his contribution
to the design and prototyping of the control-board and Mr. Jürgen
Geese for his contribution to the mechanical and electrical assembly
of the device.

%\bibliography{Ref_MAWG}

%
\end{document}